\documentclass[onecolumn]{aastex61}

\newcommand\aastex{AAS\TeX}

\accepted{21 January 2018}
\submitjournal{ApJ}

\shorttitle{\aastex\ sample article}
\shortauthors{Garc\'ia Mu\~noz}

\begin{document}

\title{On mapping exoplanet atmospheres with high-dispersion spectro-polarimetry.
Some model predictions.}

\correspondingauthor{Antonio Garc\'ia Mu\~noz}
\email{garciamunoz@astro.physik.tu-berlin.de, tonhingm@gmail.com}


\author{A. Garc\'ia Mu\~noz}
\affiliation{Zentrum f\"ur Astronomie und Astrophysik, Technische Universit\"at Berlin, \\
Hardenbergstrasse 36, D-10623, Berlin, Germany}

\begin{abstract}

Planets reflect and linearly polarize the radiation that they receive from 
their host stars. The emergent polarization is sensitive to aspects of the planet atmosphere
such as the gas composition and the occurrence of condensates and their optical properties. 
Extracting this information will represent a major step in the characterization of exoplanets.
The numerical simulations presented here show that the polarization of a spatially-unresolved exoplanet
may be detected by cross-correlating high-dispersion
linear polarization and intensity (brightness) spectra of the planet-star system.
The Doppler shift of the planet-reflected starlight 
facilitates the separation of this signal 
from the polarization introduced by the interstellar medium and the terrestrial atmosphere. 
The selection of the orbital phases and wavelengths at which to
study the planet is critical. 
An optimal choice however will partly depend on information about the atmosphere that is 
a priori unknown.
We elaborate on the cases of close-in giant exoplanets with non-uniform cloud coverage,
an outcome of recent brightness phase curve surveys from space, and
for which the hemispheres east and west of the sub-stellar point will produce 
different polarizations. 
With integration times 
on the order of hours at a 10-m telescope, 
the technique might distinguish amongst some proposed asymmetric 
cloud scenarios with fractional polarizations of 10 parts per million 
for one such planet orbiting a V-mag=5.5 host star.
Future 30-40-m telescopes equipped with high-dispersion spectro-polarimeters  
will be able to investigate the linear polarization of smaller planets orbiting fainter stars 
\textcolor{black}{and look for molecular features in their polarization spectra}.

\end{abstract}





\section{Introduction} \label{sec:intro}

Whole-disk polarization measurements of reflected sunlight 
have a long history in the remote sensing of the solar system bodies, their surfaces and
atmospheres \citep[e.g.][]{mishchenkoetal2010,kolokolovaetal2015}. 
Linear polarimetry (the focus of this work) is potentially more sensitive than brightness 
measurements to the composition, size and shape of the scattering particles. 
Both approaches complement each other in the characterization of the
condensate-gas envelope of an atmosphere  \citep{hansentravis1974}.
Venus and Titan are classical examples of how polarization has contributed to our understanding 
of atmospheres. The polarization of Venus varies strongly with wavelength and phase angle, 
and shows specific features (glory, primary rainbow) that can only be ascribed to 
spherical (and thus liquid) droplets of a narrow size distribution and
refractive index consistent with 
sulfuric acid in water \citep{hansenhovenier1974}. 
Titan strongly polarizes the incident sunlight, with fractional polarizations of up to 40-70\% at
quadrature for wavelengths from the ultraviolet to the near infrared
\citep{tomaskosmith1982,westetal1983}. 
This Rayleigh-like behavior in polarization, 
combined with evidence for non-Rayleigh-like strong forward scattering in intensity, 
is at the core of the interpretation of Titan's ubiquitous haze as forming by 
aggregation of thousands of nm-size monomers produced through photochemical
processes \citep{westsmith1991,lavvasetal2013}.

The potential and current status of polarimetry for the detection of
exoplanets and the characterization of their atmospheres and orbits 
has been discussed at length \citep[e.g.][]{seageretal2000,stametal2004,
fluriberdyugina2010,fossatietal2012,wiktorowiczstam2015}. 
Since polarimetry is a photon-starved technique, most efforts 
\citep[e.g.][]{berdyuginaetal2008,lucasetal2009,wiktorowiczetal2015,bottetal2016} 
have focused on broadband polarization from bright star systems (55 Cancri, $\tau$ Bo\"otes, HD 189733).
To date, the only claim for exoplanet polarization refers to HD 189733 b, 
a hot Jupiter orbiting a V-mag=7.7, K-type active star \citep{berdyuginaetal2008,berdyuginaetal2011},
but the detection remains contested \citep{wiktorowiczetal2015,bottetal2016}.
Exoplanet polarization will eventually transition 
into an undisputed reality, and that step will open unprecedented
possibilities for the characterization of their atmospheres. 

Broadband polarimetry ensures that a large number of photons are collected, which is 
essential to reach the required sensitivities of tens of parts per million (ppm) or better.
On the other hand, broadband polarimetry also 
requires the removal of systematics introduced by the telescope-instrument
optical system, and the subtraction of the polarization arising in the interaction 
of the starlight with the interstellar medium (ISM) or with the terrestrial atmosphere. 
In reality, the non-planet components of the  measured polarization signal 
can easily bury the planet-caused polarization.
These technical challenges are driving fast progress in polarimetry at the ppm-level,
including the definition of standard stars and the understanding of
instrumental polarization, and in the polarimetric charting of the ISM 
\citep[e.g.][]{fossatietal2007,lucasetal2009,wiktorowicz2009,baileyetal2015,cikotaetal2017,
cottonetal2016,cottonetal2017}. Ultimately, achieving sensitivities at the ppm level 
will require the joint optimization of each of these aspects.

An alternative to broadband polarimetry is the use of high-dispersion spectro-polarimetry
together with some form of cross-correlation (HDSP-CC hereafter). 
The idea was first presented by \citet{moutouetal2007}, but not developed in depth. 
The authors failed to find the polarized signal of HD 189733 b in measurements
carried out with the ESPaDOnS instrument at the 
3.6-m Canada-France-Hawaii telescope, probably because 
the signal-to-noise ratio (SNR) of their measurements was insufficient. 
In the last years, the HDS-CC technique (without polarimetry)
has consolidated as a powerful tool in the compositional and dynamical 
investigation of exoplanet atmospheres 
\citep[e.g.][]{snellenetal2010,snellenetal2014,rodleretal2012,lovisetal2017}. 
The technique benefits from
the spectral separation of the planet and the star due to their relative
Doppler shifts and from the use of a priori information about the atmospheric feature being searched for
in the form of a template spectrum. 
After a few unfruitful attempts
\citep[e.g.][]{charbonneauetal1998,colliercameronetal1999,rodleretal2013}, 
\citet{martinsetal2015} recently reported the 
detection of reflected starlight from the non-transiting hot Jupiter 51 Peg b. 
The authors inferred a geometric albedo
$A_{\rm{g}}$=0.5$\times$(1.9/($R_{\rm{p}}$/$R_{\rm{J}}$))$^2$, which is degenerate with the
undetermined planet radius (in Jovian units) $R_{\rm{p}}$/$R_{\rm{J}}$.  
This is a large albedo 
for any $R_{\rm{p}}$/$R_{\rm{J}}$ consistent with planet structure models, and makes
the planet stand out amongst more typical $A_{\rm{g}}$$<$0.2 
values for hot Jupiters \citep{angerhausenetal2015,estevesetal2015}. 
Regardless of the precise value for 51 Peg b's geometric albedo, 
the detection represents a milestone for the consolidation
of the HDS-CC technique in reflected starlight investigations.

The present work elaborates on the HDSP-CC technique in its application to
the characterization of exoplanet atmospheres.
Making its case is timely because there are a number of spectro-polarimeters
(e.g. ESPaDOnS, PEPSI, HARPSpol, SPIRou) 
\citep{donatietal2006,sniketal2008,artigauetal2014,
strassmeieretal2015} either available or 
being commissioned at 4--10 m telescopes, and 
instrument concepts under consideration for next generation space telescopes 
such as LUVOIR/POLLUX.\footnote{\url{https://asd.gsfc.nasa.gov/luvoir/}}
The paper is split as follows. 
In \S\ref{sec:patchycloud}, we introduce the scientific case selected to demonstrate 
some practical aspects of the technique. 
Section 
\S\ref{sec:photometricmodel} presents our fiducial planet and the
idealized photometric model that enables us to explore the performance of the technique, and 
\S\ref{sec:ccfperformance} discusses the outcome of the numerical simulations. 
Finally, in \S\ref{sec:discussion} we summarize the main conclusions and 
anticipate future avenues to explore.


\section{Patchy clouds on hot Jupiters} \label{sec:patchycloud}

Observations with the Kepler mission 
(photometric precisions of a few ppm over multi-year baselines, 
sensitive to 0.4--0.9 $\mu$m radiation of  
effective wavelength $\lambda_{\rm{eff}}$$\approx$0.65 $\mu$m) have enabled
the atmospheric investigation of $\sim$20 transiting exoplanets over their full orbit 
\citep[e.g.][for a review]{shporer2017}.
A conclusion drawn from such studies is that the whole-disk brightness of close-in giant planets
(hot Jupiters) may peak 
before, at or after occultation \citep{demoryetal2013,angerhausenetal2015,estevesetal2015}. 
This lack of symmetry in the brightness phase curves is explained through
an asymmetry in the planet envelopes. 
For the hotter (cooler) planets, the brightness tends to peak before (after) occultation
\citep{estevesetal2015}. 
General Circulation Model (GCM) simulations provide valuable context to understand this trend
on the basis of how the energy is transported on the global scale. 
The atmospheric super-rotation generally predicted by GCMs 
will shift the planet hot spot eastwards from the substellar point.\footnote{We 
adopt the convention that 
close-in giant exoplanets are tidally locked and follow prograde orbits. 
Hence the equivalence between west and dawn, and east and dusk.} 
At high temperatures, thermal emission from the hot spot would 
dominate the planet signal, thereby introducing a pre-occultation brightness peak. 
In contrast, cooler equilibrium temperatures would facilitate the formation
of clouds on the night side, which are transported by the eastwards jets onto the dayside
where they ultimately evaporate 
\citep{showmanetal2008,oreshenkoetal2016,parmentieretal2016,romanrauscher2017}. 
For these planets, the resulting non-uniform cloud distribution would boost the planet 
reflectivity after occultation.

Amongst the planets with post-occultation peaks, Kepler-7b has received significant attention 
\citep{demoryetal2013,garciamunozisaak2015,huetal2015,
oreshenkoetal2016, parmentieretal2016,romanrauscher2017} and become a reference for what
will be possible in the near
future with the phase curves to be delivered by missions such as CHEOPS \citep{fortieretal2014}, JWST, TESS \citep{rickeretal2015} or PLATO
\citep{raueretal2014} targeting brighter stars. 
For this reason, we will hereafter use the atmospheric structure
inferred for this planet to demonstrate the HDSP-CC technique. This choice serves also to
highlight the synergies between multi-facility investigations. 

Kepler-7b is a giant ($R_{\rm{p}}$/$R_{\rm{J}}$=1.61), low-density 
($\rho_{\rm{p}}$/$\rho_{\rm{J}}$=0.14) exoplanet orbiting a G-type star \citep{lathametal2010}, 
and exhibits an unusually high geometric albedo for a hot Jupiter 
of $\sim$0.30 \citep{demoryetal2013,garciamunozisaak2015}. 
The non-detection of occultations  at 3.6 and 4.5 $\mu$m with Spitzer
confirms that the brightness over the Kepler passband 
is dominated by reflected starlight rather than by thermal emission \citep{demoryetal2013}.
In the currently accepted scenario, 
the post-occultation peak in Kepler-7b's phase curve is caused by
clouds displaced towards the dawn terminator 
that break the east-west symmetry, boost the reflectance 
and locally mask the absorbing gas below the cloud. 

\citet{garciamunozisaak2015} proposed a scenario that synthesizes the above ideas 
while allowing to investigate cloud 
properties such as the optical thickness, horizontal extent and displacement from
the substellar point, the optical properties of the condensates, 
and the overall reflectance of the gas below the cloud. 
By solving the multiple-scattering radiative transfer problem for 
millions of configurations, each exploring a combination of the model properties,
and the $\chi^2$ comparison of the synthetic phase curves to the measurements, 
the authors identified the continuum of cloud-gas configurations that best reproduce the 
Kepler-7b brightness phase curve.
This continuum is generally characterized by a thick cloud
displaced towards the dawn terminator and resting above strongly absorbing gas. 
The cloud consists of particles that scatter the incident starlight with a single
scattering albedo close to one, thereby imparting the planet its overall large reflectance.
The equilibrium temperature of the planet, $T_{\rm{eq}}$$\sim$1630 K \citep{garciamunozisaak2015},
sets another constraint because at temperatures loosely related to $T_{\rm{eq}}$  
only a few plausible condensate compositions are consistent with 
nearly-conservative scattering. 
In the framework of Mie theory (strictly valid only for spherical particles,  
the authors further constrain the refractive index and particle size of the cloud particles. 
The inferred effective radii are: 
0.1--0.32 $\mu$m for silicate (refractive index, $n$$\approx$1.6+i10$^{-4}$); 
0.08--0.2 $\mu$m for perovskite ($n$$\approx$2.25+i10$^{-4}$); 
0.1--0.4 $\mu$m for silica ($n$$\approx$1.5+i10$^{-7}$). 
The quoted intervals bracket 4 standard deviations relative to the optimal solutions for
each condensate. 
The proposed scenarios are consistent with the expectation that the gas phase of hot Jupiter atmospheres
over the Kepler passband is strongly absorbing due to the occurrence of alkalis at the altitudes being probed
by reflected starlight photons \citep{seageretal2000,sudarskyetal2000}. 
For Kepler-7b, the hypothesized cloud, if at high altitude, would 
mute the alkali absorption, especially after occultation.
Kepler-7b's low gravity, which facilitates that small particles stay suspended
high in the atmosphere, may be key to explaining its elevated albedo \citep{sudarskyetal2000}.

\begin{figure*}
   \centering
   \includegraphics[width=9.cm]{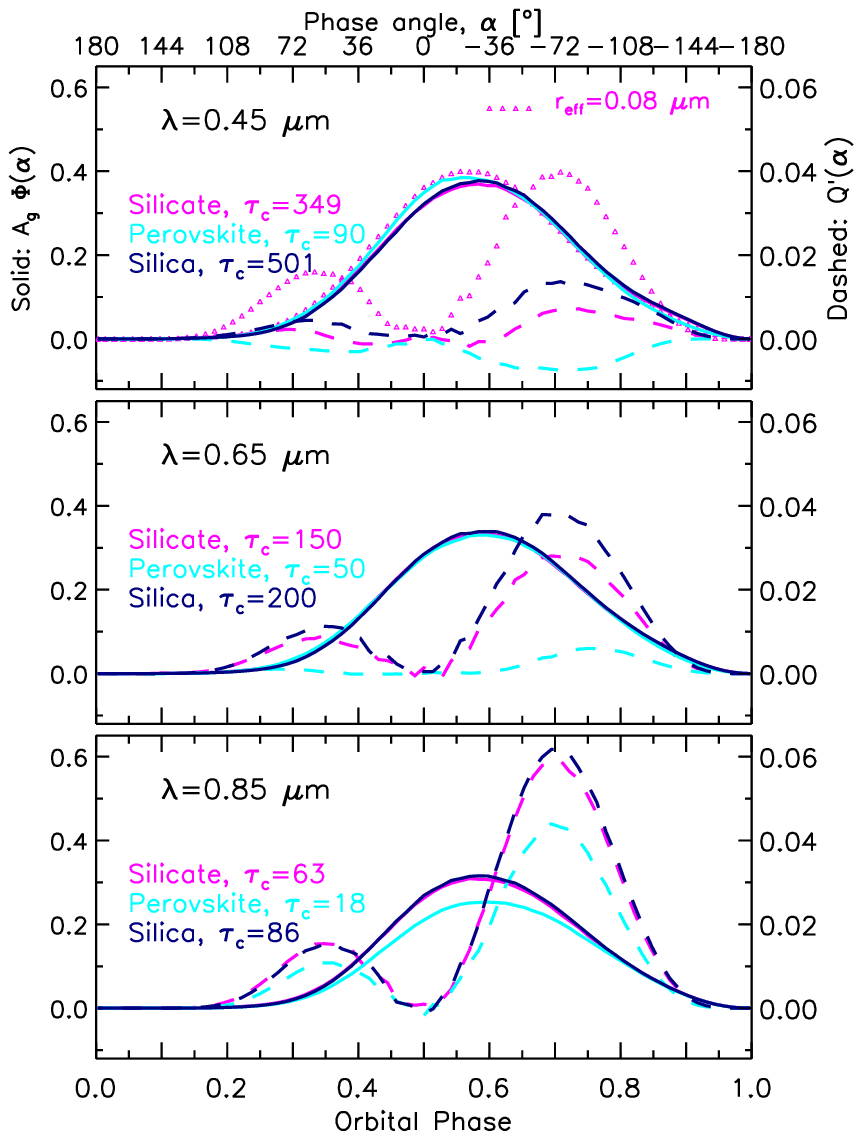}
      \caption{\label{plot2_fig} 
Brightness and polarization synthetic phase curves based on the best fits
by \citet{garciamunozisaak2015}
to the brightness measurements of Kepler-7b 
($\lambda_{\rm{eff}}$=0.65 $\mu$m) \citep{demoryetal2013}.
Three condensate compositions plausibly explain the measurements, namely: 
silicate, perovskite and silica. 
Referring to the model in \citet{garciamunozisaak2015} with a Mie treatment of the 
cloud particles, 
each cloud-gas configuration is described by five parameters: 
$\tau_{\rm{c}}$, $\sigma_{\rm{c}}$, $-\Delta\phi_{\rm{c}}$, $r_{\rm{eff}}$, and $r_{\rm{g}}$.
Our calculations here adopt $r_{\rm{g}}$=0 for all three condensates. 
Additionally, for silicate: $\sigma_c$=30$^{\circ}$; $-\Delta\phi_c$=80$^{\circ}$; $r_{\rm{eff}}$=0.13 $\mu$m.
For perovskite: $\sigma_c$=25$^{\circ}$; $-\Delta\phi_c$=55$^{\circ}$; $r_{\rm{eff}}$=0.1 $\mu$m.
For silica: $\sigma_c$=30$^{\circ}$; $-\Delta\phi_c$=85$^{\circ}$; $r_{\rm{eff}}$=0.13 $\mu$m. 
The optical thickness at the cloud center, $\tau_{\rm{c}}$, 
for each condensate and wavelength is quoted in the corresponding graph.
\textcolor{black}{$\tau_{\rm{c}}$ over the Kepler passband 
($\lambda_{\rm{eff}}$=0.65 $\mu$m) was obtained from the fit to the
measurements; at the other wavelengths, $\tau_{\rm{c}}$ is estimated on the basis of 
the ratio of particle extinction cross sections at the corresponding wavelength and at 
0.65 $\mu$m.}
     }
   \end{figure*}

Figure (\ref{plot2_fig}b) shows (solid lines) the synthetic phase curves 
for the cloud-gas configurations reproducing best the Kepler-7b measurements, 
as inferred by \citet{garciamunozisaak2015} (see caption for additional details). 
They are characterized by 
$r_{\rm{eff}}$=0.13 $\mu$m for silicate and silica, and 0.1 $\mu$m for perovskite.
The corresponding polarization phase curves are also shown (dashed lines). 
All simulations were done with a backward Monte-Carlo algorithm \citep{garciamunozmills2015}  
as described in \citet{garciamunozisaak2015}. 
The scattering matrices for each
condensate were obtained with the Mie theory model of \citet{mishchenkoetal2002}.
By construction, the three cloud-gas configurations 
produce brightness phase curves that are indistinguishable over the Kepler passband.
In contrast, 
the dashed curves of Fig. (\ref{plot2_fig}b) show that each cloud-gas configuration will
linearly polarize the incident starlight differently. 
We represent the Stokes vector for the irradiance from the planet, 
normalized to the irradiance of the (essentially) unpolarized
star, as ($R_{\rm{p}}$/$a$)$^2$$[A_{\rm{g}}\Phi, Q', U', V']$. 
The term $Q'$ used in the representation of Fig. (\ref{plot2_fig})  
is sometimes called polarized intensity \citep{buenzlischmid2009} and 
is the linear polarization equivalent of the size-normalized representation 
for the planet brightness $A_{\rm{g}}$$\Phi$. 
\textcolor{black}{
$U'$ is the linearly polarized intensity at 45$^{\circ}$ rotation and $V'$ is the 
circularly polarized intensity.}
$A_{\rm{g}}$ and $\Phi$($\alpha$) stand for the geometric albedo and the planet phase
law, respectively, and by convention $\Phi$($\alpha$=0)$\equiv$1. 
Each of the elements in brackets for the planet Stokes vector depend on both $\alpha$ and $\lambda$. 
We occasionally omit one of these dependencies to highlight the other. 
Neglecting $U'$ and $V'$, 
the fractional polarization of the stand-alone planet, i.e. if spatially resolved from its
host star, is given by the ratio $Q'$/$A_{\rm{g}}$$\Phi$. 

In Mie theory, 
the scattering matrix that goes into the radiative transfer equation 
is sensitive to the refractive index (and in turn composition) and size parameter 
($x_{\rm{eff}}$=2$\pi$$r_{\rm{eff}}$/$\lambda_{\rm{eff}}$) 
of the scatterers. If the particles are not spherical, then Mie theory does not apply and
the particles shape becomes an extra factor to consider \citep{mishchenkoetal2010}. 
As in \citet{garciamunozisaak2015}, 
we will here assume sphericity of the scattering particles as a convenient approach \textcolor{black}{
to connect the microscopic characteristics of the cloud particles, their optical 
properties and the overall appearance of the planet.} 
This pragmatic simplification 
might eventually be tested with polarization data since Mie theory results in 
well-defined predictions.

The simulations of $Q'$ in Fig. (\ref{plot2_fig}b) 
reveal the power of polarization to discriminate between cloud-gas configurations 
that produce no distinguishable behavior in brightness. 
The asymmetry in the brightness phase curves is mirrored by two distinct peaks 
in the pre- (orbital phase, OP$<$0.5; $\alpha$$>$0) and post- (OP$>$0.5; $\alpha$$<$0) 
occultation polarizations. 
Both $A_{\rm{g}}$$\Phi$ and $Q'$ peak at OP$>$0.5. 
The ratio $Q'$/$A_{\rm{g}}$$\Phi$ (not shown here) peaks nearer to quadrature, 
($|\alpha|$$\sim$90$^{\circ}$), and is larger for OP$\sim$0.25 than for OP$\sim$0.75.
Detecting the pre- and post-occultation polarization 
will help confirm or rule out some of the proposed asymmetric
cloud-gas scenarios, thereby helping map the planet.

We have predicted the planet phase curves at other wavelengths. 
Towards that end, we re-calculated the cloud optical thickness 
(which is proportional to the particles extinction cross section), and  
the scattering matrix and single scattering albedo of the particles. 
Figures (\ref{plot2_fig}a, c) present 
the phase curves at $\lambda_{\rm{eff}}$=0.45 and 0.85 $\mu$m. 
Again, minor differences in brightness can be corresponded with major
differences in polarization. 
From top to bottom, the polarization phase curves in Fig. (\ref{plot2_fig}) evolve into a
Rayleigh-like behavior, i.e. increasingly positive $Q'$/$A_{\rm{g}}$$\Phi$ ratios near quadrature \citep{buenzlischmid2009}, 
as the size parameter $x_{\rm{eff}}$ decreases. 

The simulations of Fig. (\ref{plot2_fig}) are based on one-slab 
representations of the atmosphere that do not distinguish between multiple vertical layers.
In particular, the simulations do not include a separate gas or haze layer on 
top of the column-averaged cloud. 
In brightness measurements over the Kepler passband, 
the contribution from such a layer is either embedded in the cloud properties inferred by \citet{garciamunozisaak2015} 
or is small in the near-nadir view at which the brightness phase curve is more constraining.
Indeed, a planet-wide, optically thick Rayleigh-scattering layer on top of the cloud 
would produce no post-occultation peak in Kepler-7b's brightness phase curve.
Such a gas/haze layer might affect negligibly the brightness phase curve 
(depending on its optical thickness) while
modifying substantially the polarizing properties of the atmosphere (which is dictated
by the optical depths $\le$1 as measured from the atmospheric top) \citep{buenzlischmid2009}. 
The triangle phase curves in Fig. (\ref{plot2_fig}a) illustrate this point. 
They were produced with the same combination of model parameters as for the silicate 
simulations, except that a smaller particle size $r_{\rm{eff}}$$\approx$0.08 $\mu$m
was implemented.
Although the overall planet brightness is not significantly affected, 
a decreased $x_{\rm{eff}}$ results in an increased Rayleigh-like 
polarization. The above examples show that: asymmetries in the planet phase curve 
can be corresponded by asymmetric polarizations; $Q'$ can take a range of values which are
poorly constrained by the brightness measurements. 
\textcolor{black}{In turn, measuring $Q'$ would help distinguish between some of the proposed
scenarios.}

\begin{figure*}
   \centering
   \includegraphics[width=9.cm]{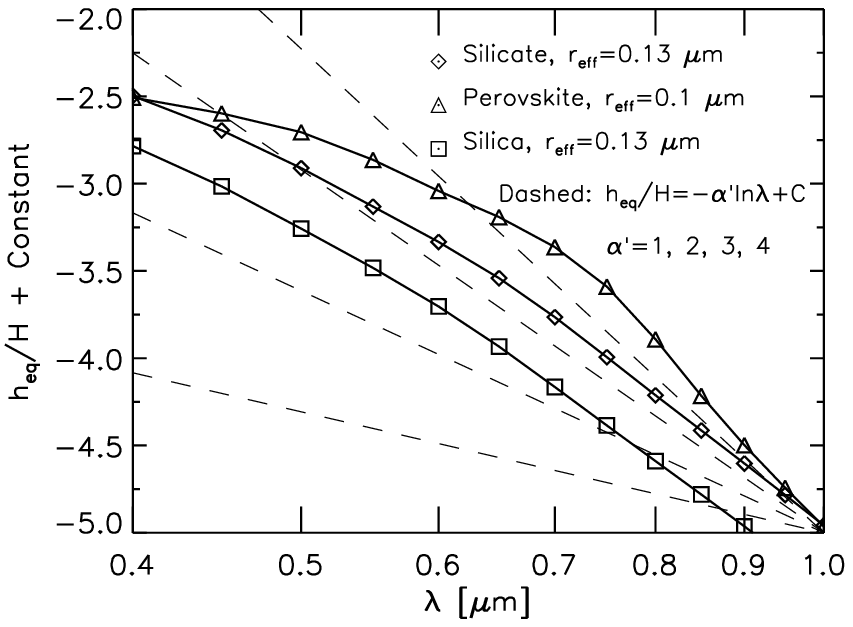}
      \caption{\label{plot_slope2_fig} 
Variation with wavelength of the atmospheric 
equivalent height normalized to the scale height. 
To within a good approximation, $h_{\rm{eq}}$/$H$ depends linearly on
$\ln{\sigma}$($\lambda$), where $\sigma$ is the extinction cross section of the 
scattering particles that interact with the starlight at the optical radius level 
 \citep{lecavelierdesetangsetal2008}. 
The solid curves represent $h_{\rm{eq}}$/$H$ for the cloud particle properties described
in Fig. (\ref{plot2_fig}). 
The dashed curves are further simplified representations of $h_{\rm{eq}}$/$H$ that 
assume $\sigma$$\propto$$\lambda^{-\alpha'}$. 
}
   \end{figure*}

If the planet was viewed in transmission, the small particles ($r_{\rm{eff}}$=0.1--0.13)
used for the simulations in Fig. (\ref{plot2_fig}) 
would produce a Rayleigh slope at short wavelengths.  
This is seen in Fig. (\ref{plot_slope2_fig}), which shows the 
equivalent height of the atmosphere normalized to the pressure scale height
for wavelengths from 0.4 to 1 $\mu$m 
according to the analytical treatment of \citet{lecavelierdesetangsetal2008}.
Thus, atmospheres with Rayleigh slopes in transmission or with 
moderately large geometric albedos do not necessarily
produce Rayleigh-like polarization phase curves across the UV-NIR spectrum.

The sensitivity of polarization to wavelength is a two-edged sword. 
Multi-wavelength polarization measurements probing a range of $x_{\rm{eff}}$ values
will likely experience a broader range of behaviors than the corresponding brightness. 
This sensitivity will translate into stronger constraints on the atmosphere. 
On the other hand, measurements averaging over a broad spectral range may inadvertently 
wash out the wavelength-dependent response of the atmosphere, and result in a 
an erroneous interpretation. 
Similarly, because $Q'$ depends on $\alpha$ in a less predictable way than 
$A_{\rm{g}}$$\Phi$, measurements over a broad range of phase angles may partly
wash out the information encoded in the polarization phase curves. 
These arguments have implications for the design of prospective observations. 
Ideally, one would carry out the polarization observations at a range of phase angles that maximize the
planet polarization, and at wavelengths that optimize both the planet and 
star responses to the HDSP-CC technique. The latter is discussed below.

\section{Photometric model} \label{sec:photometricmodel}

\begin{figure*}
   \centering
   \includegraphics[width=9.cm]{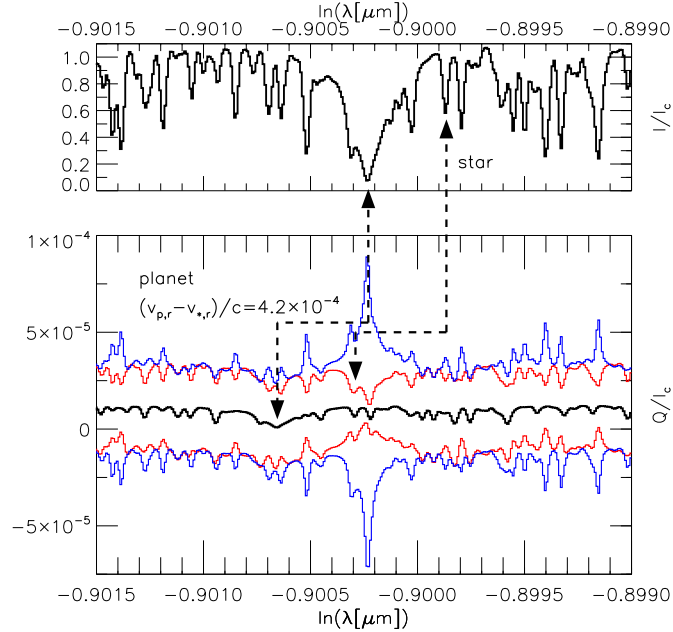}
      \caption{\label{plot1_fig} 
\textbf{Top}. Continuum-normalized stellar intensity spectrum. 
\textbf{Bottom}. After occultation, for a phase angle $\alpha$=$-72^{\circ}$, 
the planet moves towards the observer with a velocity of $\approx$126 km s$^{-1}$, 
which causes a blue shift in the planet-reflected spectrum. 
Black: 
($R_{\rm{p}}$/$a$)$^2$$Q'$$\mathcal{F}_{\star}(\lambda-\lambda_{\rm{D}})$/$\mathcal{F}_{c,\star}(\lambda)$ 
with $Q'$=0.05, which produces a contrast of $\sim$11 ppm in the continuum. 
Red: Planet polarization plus/minus $\sigma_{\rm{pn}}$($\epsilon_I(\lambda)$)/$\mathcal{F}_{c,\star}(\lambda)$. 
Blue: 
Planet polarization plus/minus 
$\sigma_{\rm{pn}}$($\epsilon_I(\lambda)$)/$\mathcal{F}_{c,\star}(\lambda)$$\times$1/$\mathcal{T}$($\lambda$). 
In both noise budget cofigurations ($b$=0 and 1), 
the noise dominates over the planet polarization. 
The cross correlation function exploits the redundancy of having multiple stellar lines over the measured
spectrum to reduce the noise below the planet signal. 
In the $b$=1 configuration, the noise at the position of strong stellar absorption bands
is significantly enlarged. 
The dashed arrows connecting the top and bottom graphs indicate the Doppler shift of the planet.  
     }
   \end{figure*}

The fiducial planet-star system that serves us to explore the HDSP-CC technique is 
constructed by combining the orbital properties and star magnitude of the 51 Peg 
system and the atmospheric properties inferred for Kepler-7b. 
In other words, 
we consider in our exercise that 51 Peg b's atmosphere is similar to the atmosphere of Kepler-7b. 
The assumption is speculative but not necessarily far fetched as both planets have
similar equilibrium temperatures and masses, and have been singled out from the general 
population of close-in giant exoplanets for their high reflectances. 
A critical aspect of 51 Peg b is that, unlike Kepler-7b, 
it orbits a very bright star, and this entails an obvious advantage to achieve the required
SNR. 
The fundamental assumptions adopted in the photometric model of our fiducial planet
are easy to re-scale for other planet-star configurations. Thus, the conclusions
drawn from the exercise go beyond the specific case represented here. 
Other simplifications in our idealized analysis include: 
we ignore the impact of telluric absorption, sky brightness corrections, cross talk 
between circular and linear polarization, and 
the specifics of the detector performance (e.g. duty cycle, read-out noise). 
Telluric absorption may become problematic in the near infrared, 
which is significantly affected by molecular oxygen and water bands. 
This may pose an additional challenge for the application of the technique on 
planets orbiting cool stars with rich spectra at these wavelengths.
If the sky brightness originates from scattered moonlight, then its spectrum and Doppler
shift are well known and can be removed \citep{donatietal2016}. 
Cross talk at the instrument level may produce spurious linear polarization \citep{bagnuloetal2009}. 
If the spectral lines of the spurious linear polarization spectrum
match the position of the lines in the stellar intensity spectrum,
the distinct Doppler-shift of the planet signal should facilitate the separation of 
each contribution.
Figure (\ref{plot1_fig}) introduces some of the concepts in the photometric model.

We adopted $a$=0.052 AU, V-mag=5.5, $M_{\rm{p}}$/$M_{\rm{J}}$=0.461,
$M_{\star}$/$M_{\sun}$=1.054, and $R_{\star}$/$R_{\sun}$=1.025 (as for the 51 Peg system\footnote{http://exoplanets.org}),
and $R_{\rm{p}}$/$R_{\rm{J}}$=1.61 (as for Kepler-7b) \citep{demoryetal2011}. 
For the stellar spectrum, we implemented Kurucz's very high-resolution solar 
spectrum\footnote{http://kurucz.harvard.edu/sun/irradiance2005/irradthuwl.dat},  
and corrected the solar irradiance $\mathcal{F}_{\odot}$ to 
$\mathcal{F}_{\star}$=$\mathcal{F}_{\odot} (\rm{AU}/d_{\star})^2$ with
$d_{\star}$=13.6 pc so that $d_{\star}/\rm{AU}=2.512^{(\rm{Vmag}+26.74)/2}$.
Without loss of generality, 
we also assumed that our fiducial planet is on a circular, 
edge-on (inclination $i$=90$^{\circ}$) orbit that traces a straight line
on the sky.  
Most transiting exoplanets are found in similar orbits and go through 
almost the full range of phase angles $\alpha$ from $-$180 to $+$180$^{\circ}$
as viewed from Earth.
\textcolor{black}{Although 51 Peg b is not occulted by its host star, we will
refer to superior conjunction as the occultation phase.}
For a given orbital period, an $i$=90$^{\circ}$ orbit maximizes the planet and star radial
velocities.
Detecting the planet's radial velocity (whether in the intensity or polarization spectra) 
provides a determination of $i$ with which to break the $M_{\rm{p}}$$\sin{(i)}$ 
degeneracy from star-only radial velocities and in turn determine the true planet mass.

We define the reference plane for polarization
to be perpendicular to the scattering plane. 
For $i$=90$^{\circ}$, the reference plane projects on the sky as a straight line
at a right angle to the planet orbit.
$Q$ is the linear polarization component in the direction parallel
($Q$$>$0) or perpendicular ($Q$$<$0) to the reference plane, and 
$U$ is the corresponding component at $\chi$=$\pm$45$^{\circ}$. 
For a planet whose hemispheres north and south of the scattering plane are not too
different, 
the emergent linear polarization at $\chi$=$\pm$45$^{\circ}$
will be minor and generally $|U'|$$\ll$$|Q'|$.
Finding the direction at which $U'$ vanishes provides a way to infer the 
orientation of the planet orbit, 
a determination that is not possible with intensity-only measurements. 
That possibility comes at the cost of requiring, at least during an initial exploration, 
measurements in two directions rotated by 45$^{\circ}$ 
(or four to remove to first approximation the effect of instrumental polarization 
\citep{bagnuloetal2009}). 
For simplicity, 
we will assume in our calculations that the orientation of the planet orbit on the 
sky is known and that the corresponding Stokes element $Q$ is directly measurable with a
single exposure and without additional rotations. The estimated exposure times
can be properly re-scaled if needed to relax these assumptions.

Defined the reference plane, 
the intensity and linear polarization observables of the planet-star system are 
represented in the photometric model through the expressions:
\begin{equation}
m_{I}(\lambda; \alpha)=
\left(\frac{R_{\rm{p}}}{a} \right)^2 A_{\rm{g}}(\lambda) \Phi(\lambda; \alpha)
\mathcal{F}_{\star}(\lambda-\delta\lambda_{\rm{D}}) + \mathcal{F}_{\star}(\lambda) + 
\epsilon_I(\lambda) 
\label{mi_eq}
\end{equation}
\begin{equation}
m_{Q}(\lambda; \alpha)=\left(\frac{R_{\rm{p}}}{a} \right)^2 Q'(\lambda; \alpha) 
\mathcal{F}_{\star}(\lambda-\delta\lambda_{\rm{D}}) + m^{\rm{ISM}}_{Q}(\lambda) + \epsilon_Q(\lambda)
\label{mq_eq}
\end{equation}
We omit the corresponding $m_{U}$ because if $|U'|$$\ll$$|Q'|$ it provides
no information about the planet. 
We also omit the $m_{V}$ term corresponding to circular polarization because cancelation
between the north and southern hemispheres will expectedly drive its value small
\citep[e.g.][]{garciamunoz2014,wiktorowiczstam2015}.
Both $m_{I}$ and $m_{Q}$ depend on wavelength $\lambda$ and on the star-planet-observer
phase angle $\alpha$. Here, $\lambda$ is defined on the rest reference frame of the star, 
and $\delta\lambda_{\rm{D}}$ is the relative Doppler shift of the planet, i.e.
$\delta\lambda_{\rm{D}}/\lambda=(v_{\star,r}-v_{p,r})/c$, where 
$v_{p,r}-v_{\star,r}$ is the planet-star relative radial velocity ($>$0 if towards the observer) 
and $c$ the speed of
light.
Typically, $m_{I}$ will be dominated by the stellar flux $\mathcal{F}_{\star}(\lambda)$.  
In our photometric model, 
$\epsilon_I(\lambda)$ represents both stellar photon noise and
uncorrected systematics, and satisfies
$\epsilon_I(\lambda)$$\ll$$\mathcal{F}_{\star}(\lambda)$. 
The planet contribution to $m_{I}$ is
scaled by $\mathcal{F}_{\star}(\lambda-\delta\lambda_{\rm{D}})$,  
and considers the Doppler shift in the starlight spectrum reflected by the planet. 
It is tacitly assumed that the planet is tidally locked and that the lines in the 
measured stellar spectrum and in the planet-reflected spectrum are comparably broad.

In the observable $m_{Q}$ of Eq. (\ref{mq_eq}), 
$Q'$ contains the information on how the planet atmosphere polarizes the incident starlight. 
$Q'$ will generally vary with both wavelength and phase angle in ways that are
unpredictable without the detailed knowledge of the atmosphere (see Fig \ref{plot2_fig}).
$\epsilon_Q(\lambda)$ is analogous to $\epsilon_I(\lambda)$, and 
we adopt for their photon noise standard deviations 
$\sigma_{\rm{pn}}$($\epsilon_I(\lambda)$)=$\sigma_{\rm{pn}}$($\epsilon_Q(\lambda)$)=($\mathcal{F}_{\star}(\lambda)$)$^{1/2}$.
We will further assume that $\epsilon_Q(\lambda)$ (and $\epsilon_I(\lambda)$, although the
latter is not discussed) is deteriorated
with respect to photon noise at wavelengths near the core of strong stellar lines. 
Accordingly, we will describe 
$\epsilon_Q(\lambda)$ by means of a Gaussian random distribution with standard deviation
$\sigma_{\rm{pn+sys}}$($\epsilon_Q(\lambda)$)=$\sigma_{\rm{pn}}$($\epsilon_Q(\lambda)$)$\times$1/$\mathcal{T}^b$($\lambda$).
$\mathcal{T}$($\lambda$)=$\mathcal{F}_{\star}$($\lambda$)/$\mathcal{F}_{\rm{c},\star}$($\lambda$)
is the continuum-normalized stellar spectrum, and
$\mathcal{F}_{\rm{c},\star}$($\lambda$) a low-order polynomial representation of the 
stellar continuum. $\mathcal{T}$($\lambda$) becomes small 
at the core of stellar lines and nearly one in the continuum. 
With $b$$>$0 we adjust the potential impact of e.g. an imperfect wavelength solution 
in the extraction of $m_Q$ but also of other effects such as cross-talk in the instrument 
that are not explicitly represented in the photometric model.
We refer to these effects as systematics, even though the term may not be fully representative
of its intended meaning. 
$b$ ranges from $b$=0 (no impact, only photon noise) to $b$$\sim$1
(major impact, at wavelengths matching 
strong stellar lines systematics dominate the error budget).

Three additional processes unrelated to the planet also contribute to $m_{Q}$. 
The first of them refers to the intrinsic polarization of the star, 
which occurs at the continuum and band level.
For the Sun, the so-called second solar spectrum differs 
from the corresponding intensity spectrum, a fact that presents
opportunities to investigate the solar atmosphere \citep{stenflokeller1997}.
Upper limits on the broadband whole-disk fractional polarization of the Sun are
8$\times$10$^{-7}$ and 2$\times$10$^{-7}$ in B and V band, respectively \citep{kempetal1987}. 
Such polarizations are 
likely smaller than the  
polarizations of close-in giant planets unless their atmospheres are very depolarizing.
The second effect refers to the polarization of starlight by aligned, 
non-spherical dust grains in the ISM. Its broadband impact on the light 
reaching the telescope as a function 
of wavelength is often described in terms of the Serkowski law \citep{serkowski1973}:
\begin{equation}
\frac{p^{\rm{ISM}}(\lambda)}{p^{\rm{ISM}}(\lambda_{\rm{max}})}\approx\exp[ -1.15 \ln^2 ( \frac{\lambda_{\rm{max}}}{\lambda} ) ].
\label{serkowski_eq}
\end{equation}
Here, $p^{\rm{ISM}}(\lambda_{\rm{max}})$ stands for 
the maximum fractional polarization, occurring at $\lambda_{\rm{max}}$. 
$p^{\rm{ISM}}(\lambda)$ depends on the direction towards the star and its distance from Earth. 
Within the solar vicinity, 
this variation is typically $\leq$2 ppm pc$^{-1}$ \citep{marshalletal2016}, 
which may nevertheless suffice to dominate over the planet contribution. 
In our photometric model we will adopt 
$m^{\rm{ISM}}_{Q}(\lambda)$=$p^{\rm{ISM}}(\lambda)$$\mathcal{F}_{\star}(\lambda)$ and
$p^{\rm{ISM}}$($\lambda_{\rm{max}}$=0.55 $\mu$m)=30 ppm. 
A key point in this simple description of $m^{\rm{ISM}}_{Q}(\lambda)$ is that, 
if $p^{\rm{ISM}}(\lambda)$ is a smooth function of wavelength as described by Eq. (\ref{serkowski_eq}), 
its contribution will peak in the CC function (CCF) at the velocity of the star.
In that case,  
it might be possible to separate it from the planet polarization 
thanks to their relative Doppler shift, as discussed below. 
The third effect, physically related to the previous one, 
is the polarization of starlight introduced by the terrestrial atmosphere. 
In conditions of elevated dust content overhead the observation site, 
the grains suspended in the atmosphere also 
polarize the incident starlight to levels of up to tens of ppm \citep{baileyetal2008}. 
If uncorrected for, this telluric polarization will also bury the planet signal. 
In our photometric model, we will assimilate this latter effect into 
$m^{\rm{ISM}}_{Q}(\lambda)$, thereby assuming that its wavelength dependence is also smooth.

For our fiducial planet, $(R_{\rm{p}}/a)^2$$\sim$220 ppm, and if $Q'$$\sim$0.01--0.05, 
the fractional polarizations attributable to the planet in Eq. (\ref{mq_eq}),  
$Q'$$\times$$(R_{\rm{p}}/a)^2$,  are in the 2--11 ppm range.
Such minute signals
require between 7.4$\times$10$^{10}$ and 2.2$\times$10$^{12}$ photons
to reduce the standard deviation of the stellar photon noise to 1/3 of the signal. 
This estimate exposes one of the pitfalls of exoplanet polarimetry: 
that it is more photon-starved than photometry because $Q'$ can be significantly
less than $A_{\rm{g}}$$\Phi$($\alpha$).
For reference, past broadband measurements for HD 189733 b have claimed 
upper limits on the planet polarization of 
$\sim$30--60 ppm \citep{wiktorowiczetal2015,bottetal2016}.

For $m_{I}$ and $m_{Q}$ to represent observables at the telescope, they must incorporate
a description of the optical system and the exposure time of the observations. 
If $\eta$ stands for the end-to-end throughput of the entire optical system, 
$D$ the primary mirror diameter of the telescope, and $t_{\rm{expo}}$ the integration time, 
 the number of counts at the detector per spectral bin and the corresponding photon 
noise are obtained 
by replacing $\mathcal{F}_{\star}$$\rightarrow$$\mathcal{F}_{\star} \eta \pi (D/2)^2 t_{\rm{expo}}$
in Eqs. (\ref{mi_eq})--(\ref{mq_eq}).
After this, the units of $m_{I}$ and $m_{Q}$ are counts per spectral bin. 
For our reference calculations, we adopt $\eta$=0.1 and $D$=10 m, and 
explore the impact of $t_{\rm{expo}}$ on the SNR. 
The numerical choice of $\eta$ is motivated by the performance of existing
high-dispersion spectro-polarimeters.  
Our nominal spectral grid adopts a resolving power $R$=60,000  
and that each resolution element is sampled by two spectral bins. 
The grid is evenly spaced in $x$=$\ln{\lambda}$ and, when covering the 
0.4--1 $\mu$m range, it comprises a total of $N$=$2R \ln{(1.0/0.4)}\approx$110,000 
bins each with $\Delta x=\Delta \lambda_i / \lambda_i=1/2R\approx 8.3 \times 10^{-6}$. 
At maximum elongation, the relative planet-to-star radial velocity 
$v_{\rm{p}}$$-$$v_{\star}$$\approx$$v_{\rm{p}}$$\approx$ 134 km s$^{-1}$ 
introduces a Doppler-shift in the planet-reflected spectrum 
$\delta\lambda_{\rm{D}}/\lambda$$\sim$4.5$\times$10$^{-4}$ or up to 55 spectral bins
for $R$=60,000. 

\section{The CC function. Numerical simulations} \label{sec:ccfperformance}

\citet{sparksford2002} have described the fundamentals of 
the HDS-CC technique for the detection and atmospheric
characterization of exoplanets. 
Many of the principles introduced in that work are directly applicable here, and
we generally follow their treatment. 
Focusing on polarization, we set out to extract 
$(R_{\rm{p}}/a)^2$$Q'$ from $m_{Q}$ (Eq. \ref{mq_eq}), the latter being the quantity
determined at the telescope. 
For cross-correlation purposes, it is convenient to remove the low-frequency variations
with wavelength in $m_{Q}$ by introducing $\mu_{Q}$=$m_{Q}$/$\mathcal{F}_{\rm{c},\star}$.
$Q'$ (and therefore $\mu_{Q}$) is modulated in both wavelength and phase angle in a way
that depends on the properties of the planet atmosphere being investigated.
For the science cases of Fig. (\ref{plot2_fig}), 
$Q'$ \textcolor{black}{at a phase angle $\alpha$=$-$72$^{\circ}$}
varies by factors of up to $\sim$6 when going from a wavelength of
0.4 to 1 $\mu$m. 

The success of the CC relies on proposing an accurate template for
the planet signal hidden in the noise. 
Equation (\ref{mq_eq}) shows that the planet polarization term is essentially a scaled-down, 
Doppler-shifted version of the stellar intensity spectrum, possibly modulated in wavelength by $Q'$. 
The continuum-normalized spectrum of the star $\mathcal{T}$($\lambda$)
is thus an appropriate template to extract the planet polarization contribution from $m_{\rm{Q}}$. 
In discretized form, we have
$\mathcal{T}_i$=$\mathcal{T}$($\lambda_i$), with index $i$ running over the $N$ bins in the spectral grid.
We also define $\hat{t}_i$=$\mathcal{T}_i/\sum \mathcal{T}_i-1/N$ and $t'_i$=$N \mathcal{T}_i/\sum \mathcal{T}_i$, which
satisfy $<$$\hat{t}$$>$=$\sum \hat{t}_i/N\equiv0$ and $<$$t'$$>$=$\sum t'_i/N\equiv1$
\citep{sparksford2002}. 
The standard deviation of $t'$ over a specified wavelength range, $\sigma(t')$, is a measure
of how much structure exists in the template. 
In the current context, structure is equivalent to the spectrum having 
numerous strong lines. A small number of lines or lines that are shallow is equivalent to
a spectrum without much structure. 
The concept is important because 
the more structured the template is, the less likely it is that a peak is produced in 
the CCF by random noise as this should cancel out in the process of adding
up the contribution from many spectral bins. 
\citet{sparksford2002} provide theoretical arguments showing that the achievable 
SNR from the CC is indeed proportional to (and thus partly limited by) $\sigma(t')$. 
In practice, this means that the photons received over spectral regions with little structure 
(small $\sigma(t')$) are essentially useless to the effects of the CC. 
$\sigma(t')$ is straightforward to quantify from the stellar intensity spectrum and, 
leaving aside considerations on the planet polarization, it
provides a-priori insight into which spectral regions are better behaved for application of the CC.
We have done so for the Sun, and 
obtained $\sigma(t')$=0.24, 0.14, 0.072, 0.052, 0.068, 0.041 for wavelengths in the 
ranges (in $\mu$m): 0.4--0.5, 0.5--0.6, 0.6--0.7, 0.7--0.8, 0.8--0.9 and 0.9--1.0, 
respectively. All other factors being equal, 
the calculated $\sigma(t')$ tells us that the HDSP-CC technique will use the
stellar photons of solar-like stars a few times more efficiently at blue wavelengths than 
in the near infrared. This disparate efficiency is dictated by the structure of the
stellar spectrum, richer at the shorter wavelengths, 
and not by the integrated number of photons from each spectral interval, 
which is comparable in all cases.

In its simplest form, the CCF is formed by the operation:
\begin{equation}
C_j=\sum_i \mu_{Q,i} \hat{t}_{i-j} 
\label{CCF_eq}
\end{equation}
For our nominal spectral grid, a shift by one bin is equivalent to a velocity displacement 
$v$=$(\Delta \lambda_i / \lambda_i) c$$\approx$2.5 km s$^{-1}$.

\begin{figure*}
   \centering
   \includegraphics[width=9.cm]{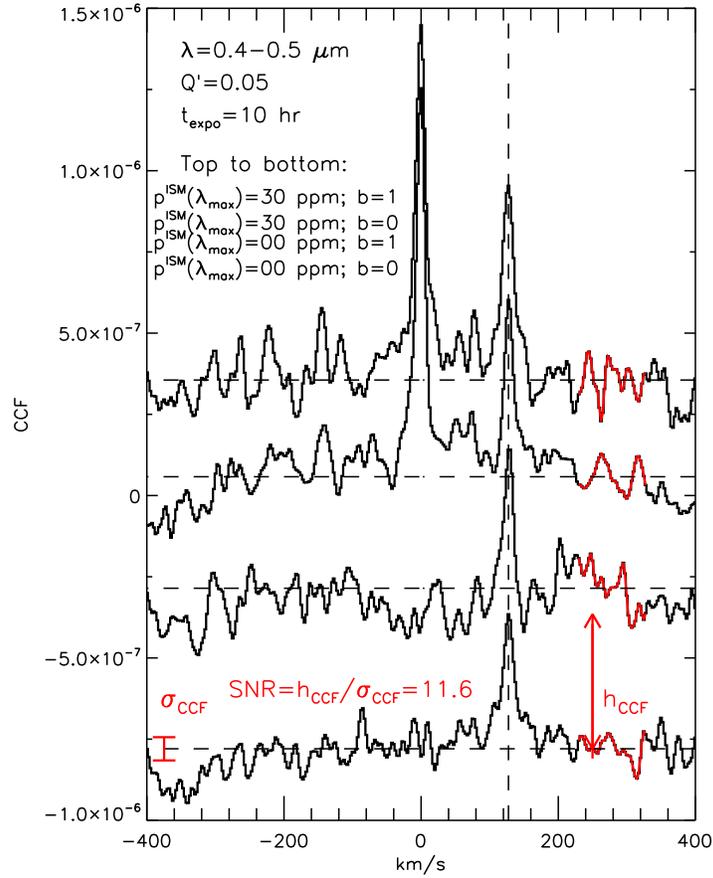}
      \caption{\label{plot3_fig} 
Examples of cross correlation based on Eq. (\ref{CCF_eq}) for the indicated configurations. 
The SNR is defined as the ratio of the CC peak height (subtracting the pedestal value
estimated far from the peak) (right arrow, lowermost graph) 
over the standard deviation of the CC in the same continuum (red bar, lowermost graph). 
To estimate the continuum, we take a region between $+$100 and $+$200 
km s$^{-1}$ from the planet (shown in red in the CC representation). 
}
   \end{figure*}

Figure (\ref{plot3_fig}) shows the CCF for $Q'$=0.05, $t_{\rm{expo}}$=10 h, 
the spectral interval 0.4--0.5 $\mu$m, 
and our reference telescope-instrument set-up. 
Based on Fig. (\ref{plot2_fig}), it is assumed that $\alpha$=$-72$$^{\circ}$, 
which sets the planet-star relative radial velocity to 
$\sim$127 km s$^{-1}$.
The four CCFs explore four combinations of the prescribed $m^{\rm{ISM}}_{Q}$ 
($p^{\rm{ISM}}$($\lambda_{\rm{max}}$=0.55 $\mu$m)=0 and 30 ppm; Eq. \ref{serkowski_eq})
and the contribution of systematics to the overall noise budget ($b$=0 and 1). 
The SNR is defined by the ratio of the peak-to-wing height of the CCF and the standard
deviation in a region of the CCF far enough from the peak, 
i.e. SNR=$h_{\rm{CCF}}$/$\sigma_{\rm{CCF}}$. 
The SNRs estimated in Fig. (\ref{plot3_fig}) are comparable and in the range 7--13. 
The specific SNR for any one example is partly dictated by 
random noise and varies between simulations. 
A main conclusion from Fig. (\ref{plot3_fig}) is that the planet peak in the CCF
 is well separated from the stellar peak 
(centered at zero velocity), and that it is possible to unambiguously detect the planet
signal with hours of exposure time, depending on $Q'$ and the other assumed properties of 
the photometric model.  
We have not conducted an assessment of false alarm probabilities, 
but a visual inspection over numerous simulations suggests that a robust detection of the
planet peak requires SNR$\ge$5--6.

\begin{figure*}
   \centering
   \includegraphics[width=9.cm]{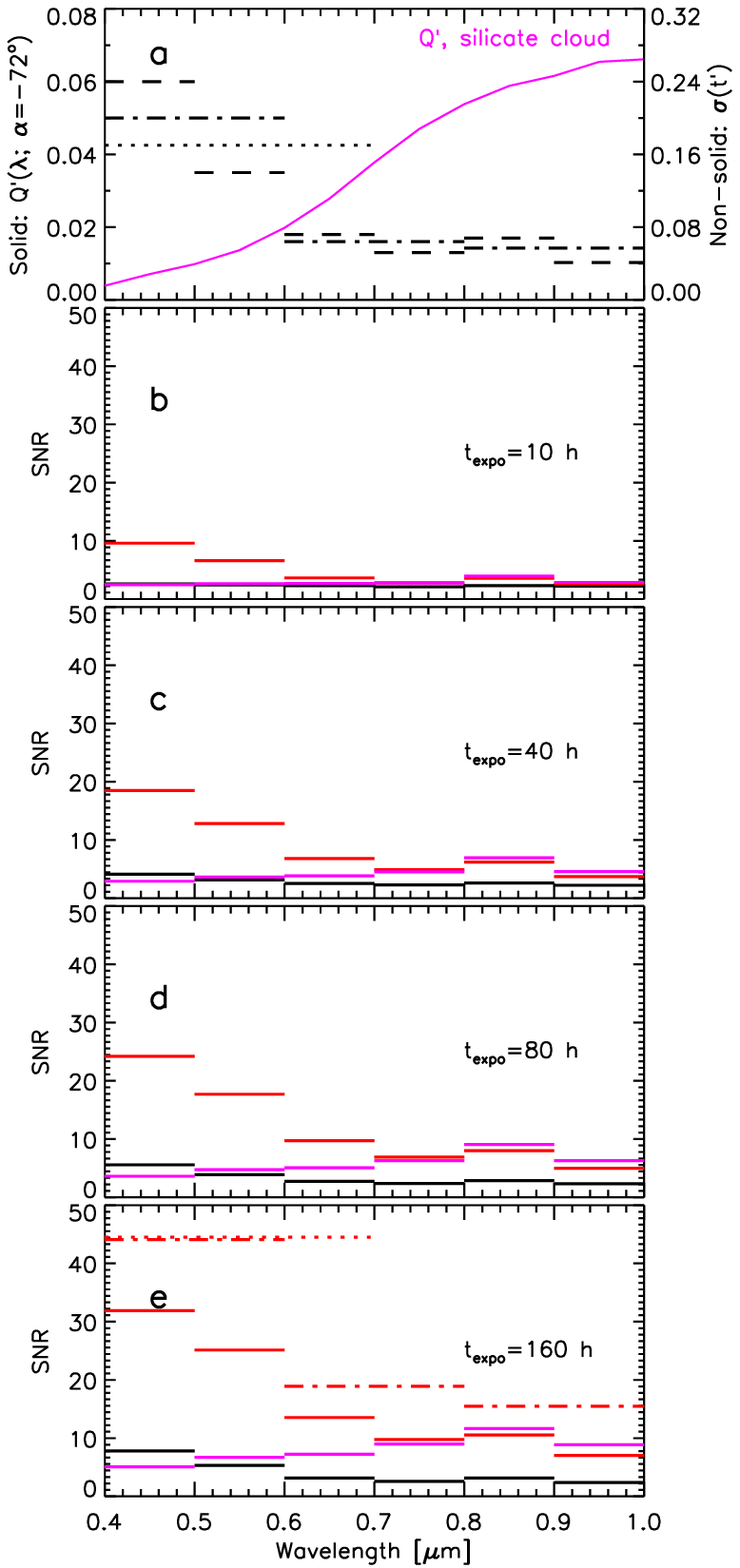}
      \caption{\label{plot4_fig} 
      \textbf{a.} Solid: Polarization element $Q'$ as a function of wavelength in the
      post-occultation configuration $\alpha$=$-$72$^{\circ}$ 
      for silicate cloud particles (Fig. \ref{plot2_fig}). 
      The planet polarizes more efficiently at the longer wavelengths. 
      \textcolor{black}{The cases $Q'$=0.01 and 0.05, independent of wavelength, are
      specifically developed in the b--e graphs with black and red colors, respectively.}
      Non-solid: Standard deviation of the normalized form of the stellar spectrum $t'$ over
      the indicated wavelengths. The stellar spectrum is significantly more structured
      at the shorter wavelengths. \textcolor{black}{Considering wavelength ranges of different
      size (from 0.1 up to 0.3 $\mu$m) reveals that a larger spectral coverage does not necessarily 
      result in better SNRs. The magnitude of $Q'$ and its wavelength dependence, the
      total number of collected photons and the structure of the stellar spectrum over the
      specified wavelength range, as quantified by
      $\sigma$($t'$), can partly compensate (see text).}
      \textbf{b--e.} SNR calculated as described in the text for: 
      $Q'$=0.01 (black), $Q'$=0.05 (red), and the wavelength-dependent $Q'$ 
      described in the top graph for silicate cloud particles. 
      Each graph refers to an exposure time from 10 to 160 h in our reference scenario. 
      We adopted for the SNR calculations 
      $p^{\rm{ISM}}$($\lambda_{\rm{max}}$=0.55 $\mu$m)=0 and $b$=0 (Fig. \ref{plot3_fig}). 
      The quoted SNRs are median values based on a few hundred random realizations of the
      noise terms in $m_Q$ (Eq. \ref{mq_eq}).
     }
   \end{figure*}

We have estimated the SNR for other configurations relevant to the science case 
introduced in \S\ref{sec:patchycloud}. Based on the above, we safely adopted 
$p^{\rm{ISM}}$($\lambda_{\rm{max}}$=0.55 $\mu$m)=0 and $b$=0.
Motivated by the phase curves in Fig. (\ref{plot2_fig}), 
we again considered an orbital position with 
$\alpha$=$-72$$^{\circ}$ and three possible behaviors for $Q'$($\lambda$). 
$Q'$($\lambda$)=0.01 is a pessimistic scenario with a weak planet polarization 
of ($R_{\rm{p}}/a$)$^2$$Q'$$\sim$2 ppm. 
$Q'$($\lambda$)=0.05, and independent of wavelength, is a plausible representation for
an atmosphere with a top layer that is moderately polarizing (Fig. \ref{plot2_fig}a, 
triangles), and results in ($R_{\rm{p}}/a$)$^2$$Q'$$\sim$11 ppm. 
Finally, we considered a $Q'$($\lambda$) that varies with wavelength 
as in the silicate cloud case of Fig. (\ref{plot2_fig}a--c). 
In this case, the planet polarizes weakly ($Q'$$<$0.01) in the blue but
moderately ($Q'$$>$0.06) in the near infrared, which also means that 
$Q'$ and $\sigma$($t'$) (for the solar spectrum) vary in the reverse orders with 
wavelength. The $Q'$($\lambda$) behavior for this latter case is shown in Fig. (\ref{plot4_fig}a).

The results of the SNR analysis are summarized in Fig. (\ref{plot4_fig}b--e) 
for the reference telescope-instrument set-up ($\eta$=0.1, $D$=10 m, $R$=60,000). 
Black, red and magenta colors are specific to the $Q'$=0.01, =0.05 and silicate scenarios 
described above, respectively.
From top to bottom, each graph refers to increasing exposure times. 
According to our photometric model, 
an exposure of 160 h at a 10-m telescope is equivalent to 
a 10-h exposure from a future 40-m class telescope. 
The width of the horizontal bars in the SNR representations conveys 
the assumed wavelength range for the observations. 

At least four factors must be considered when proposing a wavelength range for observations. 
The first one is that in general a narrow interval (say, 0.1 $\mu$m) will ensure that 
$Q'$ is nearly constant over it and thefore $\hat{t}$ remains a better template for the
planet polarization term in $\mu_{Q}$. 
The second one is that extending the observations to wavelengths at which $Q'$ is
 small may hamper rather than help the detection.
The third one is that a larger interval means that more photons are collected overall,
although this does not necessarily translate into a better SNR. 
Finally, the fourth factor refers to the level of structure
in the template, as quantified by $\sigma$($t'$), over the specific spectral region. 
If the spectral region shows little structure (small $\sigma$($t'$)), 
it will help little (or nothing) towards detecting the planet
polarization. Some of these factors are potentially in mutual conflict.
And for those depending on the magnitude and shape of $Q'$ there is limited a priori information 
that can be used to optimize the selection of wavelengths. 
Obviously, if the observations are made over a broad wavelength range, there is always the 
option of forming smaller spectral ranges in the post-observation analysis. 
The cases investigated in Fig. (\ref{plot4_fig}) demonstrate some of these possibilities
\textcolor{black}{by considering a variety of wavelength ranges over the spectrum.
They are (in $\mu$m): 0.4--0.5, 0.5--0.6, 0.7--0.8, 0.8--0.9, 0.9--1 (dashed); 
0.4--0.6, 0.6--0.8, 0.8-1 (dashed-dotted); 0.4--0.7 (dotted).}

The case with $Q'$=0.01 (black bars) is always difficult to 
detect above our tentative detection threshold of SNR=6. 
A convincing detection is 
possible though for the longest exposure time ($t_{\rm{expo}}$=160 h) 
and shortest wavelengths (0.4--0.5, and 0.5--0.6 $\mu$m). 
Unsurprisingly, 
the configuration with $Q'$=0.05 (red bars) is much less challenging, and would be 
detectable with less than $t_{\rm{expo}}$=10 h at short wavelengths. 
At the longer wavelengths, the drop in $\sigma$($t'$) however implies that much longer integration
times are needed. 
The dependence of the SNR on $\sigma$($t'$) is clearly seen 
in the configuration with $Q'$=0.05 and $t_{\rm{expo}}$=160 h (bottom graph, red bars, 
and compare with top graph, black bars). 
In this case, the SNR scales almost exactly with $\sigma$($t'$) and
makes the SNR at short wavelengths larger by a factor of a few than in the near infrared.
The bottom graph reveals also the effect of enlarging the wavelength interval. 
The SNR improvement when integrating over 0.4--0.6 $\mu$m with respect to the
corresponding SNRs at 0.4--0.5 or 0.5--0.6 $\mu$m is mainly driven by the increased
number of photons in the larger interval. However, extending the wavelength interval to
0.4--0.7 $\mu$m is compensated for by decreasing $\sigma$($t'$) values and does not
result in an improved overall SNR. 
The simulations with $Q'$=0.01 and 0.05
show also that the estimated SNR increases as the magnitude of $Q'$
or the exposure time increases, but in a manner slower than linear. 

Finally, the case of $Q'$ consistent with the silicate cloud 
(magenta) shows the compensating effects of a template
that becomes less structured towards the long wavelengths while the 
planet polarization signal becomes stronger. 
The simulations show that the resulting SNR peaks at wavelengths from
0.8 to 0.9 $\mu$m, and that this could be detected in $\sim$40 h of integration time. 
The confirmation of the silicate cloud configuration at all wavelengths from 0.4 to 1 $\mu$m 
would take $\sim$160 h at each 0.1-$\mu$m spectral range.

Finally, a few words on the efficiency with which the HDSP-CC technique utilizes the collected light. 
We integrated the first term on the right hand side of Eq. (\ref{mq_eq}) 
over the spectral direction to calculate the number of counts 
at the detector originating from planet polarization, and called this amount $s_{p,Q}$.
We similarly added up the counts arriving directly from the star, 
and called it $S_{\star}$. SNR$_{\rm{lim}}$=$s_{p,Q}$/$\sqrt{S_{\star}}$ is the limiting
signal-to-noise ratio that might be achieved if stellar photon noise dominates the noise budget. 
The comparison of the CCF-based SNR and SNR$_{\rm{lim}}$ 
reveals that SNR/SNR$_{\rm{lim}}$$\sim$$\sigma$($t'$) \citep{sparksford2002}.
The reason for this reduced efficiency of the HDSP-CC technique vs. broadband polarimetry
is that the HDSP-CC technique benefits only from photons in 
spectral regions that are highly structured. 
The photons collected in regions with little structure are essentially useless. 
Unlike broadband polarimetry, however, the HDSP-CC offers an intrinsic 
way to discriminate between polarization arising at the planet 
and other sources of polarization provided that the different polarization sources 
are Doppler-shifted.
Finally, 
we have explored the SNRs estimated in Fig. (\ref{plot4_fig}) at resolving powers 
as low as $R$=10,000, and found that on average the corresponding SNRs are reduced by 
less than 20--30$\%$. Thus, somewhat lower resolving powers do not significantly affect
the power of the HDSP-CC technique.

\section{Discussion and summary} \label{sec:discussion}

We have discussed the HDSP-CC technique and its application to the investigation of 
exoplanet polarization. The effort is motivated by the potential of polarimetry to 
offer insight into exoplanet atmospheres that is not possible with other techniques, and
by the present and near-future availability of telescopes and instruments with which to
carry out such observations. 
We presented plausible multi-wavelength polarization simulations of the close-in giant
exoplanet Kepler-7b for which broadband, optical brightness phase curves have already provided 
partial characterization of its atmosphere. 
Assuming that the atmospheric properties of
the non-transiting, bright-host hot Jupiter 51 Peg b are not too different from those
simulated for Kepler-7b, we show that the polarization of 51 Peg b might be 
an accessible target for a dedicated observational effort. 
Larger collecting areas will bring the possibility to target fainter
host stars and smaller planets, and therefore the proposed technique may develop
its full potential with the advent of 30--40-m class telescopes in the next decade. 
The characterization of exoplanet atmospheres with the HDSP-CC technique thus represents 
a valuable science case for the development of such instruments.
Particularly interesting is the future possibility of targeting planets 
within their habitable zones to investigate through polarization their gas composition 
\citep{stametal2004} or the occurrence of water clouds on them \citep{bailey2007,garciamunoz2014}. 
A related instrument concept that may one day fly is LUVOIR/POLLUX, focused on the
ultraviolet. Short wavelengths represent a gain in the structure of the stellar
spectrum and therefore in the efficiency of the HDSP-CC technique provided that the 
planet is sufficiently reflective and polarizing at such wavelengths. 
A detailed study will help establish the relevant trade-offs in the design
of reflected starlight polarization observations at ultraviolet wavelengths.

In our idealized analysis, 
we did not address the inverse problem of inferring the planet polarization from 
simulated observations. Such an exercise will depend on the specifics of each facility,
and might have to consider a range of $Q'$($\lambda; \alpha$) representations with the
goal of maximizing a figure of merit (e.g. the CCF peak or the SNR). \\

\textcolor{black}{In principle, the HDSP-CC technique can also be applied to the detection 
of atomic and molecular absorption in polarization spectra. 
\citet{sparksford2002} and \citet{lovisetal2017} have discussed some of the practicalities
for the detection of absorption features in brightness spectra. 
A possible strategy involves the construction of templates that contain the
contribution from various amounts of the gas whose existence is suspected. 
In terms of the photometric model, this
means that the term $Q'(\lambda; \alpha)$ in Eq. (\ref{mq_eq}) incorporates the 
continuum reflection by the gas (Rayleigh scattering) and cloud particles (as done here), 
together with the effect of narrowband absorption by atoms and molecules (not done here). 
The template spectrum that optimizes the CCF on the basis of a selected figure of merit 
would point to the most plausible atmospheric configuration. 
Molecules with complex spectra at high resolution
will likely provide the needed structure to make the molecule stand out in the CCF
\citep{rodlerlopezmorales2014}. For close-in giant exoplanets such as 51 Peg b, 
H$_2$O and TiO may be plausible targets for 
such searches \citep{singetal2016,sedaghatietal2017}. 
Further into the future, the search for molecular absorption might also target gases associated with
potential 
habitability conditions on the planet (e.g. O$_2$, H$_2$O or CH$_4$), all of which 
absorb at visible-NIR wavelengths. 
Indeed, the simulations by \citet{stam2008} and \citet{emdeetal2017} for various 
Earth-like atmospheric configurations show that some of these molecules introduce significant structure
in the $Q$ spectrum of polarized intensity.
Whether it is more convenient to detect the molecules in the brightness or in the
polarization spectrum will partly depend on the depth of the
absorption lines in the corresponding spectra and on the total number of photons that
can be collected in each case. It will also be interesting to find out what additional 
information could be inferred from the simultaneous detection of a molecule in both
the brightness and polarization spectra.} 

\textcolor{black}{Finally, 
\citet{lovisetal2017} have shown that a 10-m telescope with the combined capability to
partly reject the stellar glare at the planet's location 
through a high-contrast imager and to disperse the collected
light at high spectral resolution can constrain the albedo of Proxima b 
\citep{anglada-escudeetal2016} and detect the possible occurrence of O$_2$, H$_2$O or CH$_4$
in its brightness reflected spectrum in about 60 nights of telescope time. 
Provided that the planet polarizes efficiently, as expected for some Earth-like atmospheric configurations
\citep{stam2008,garciamunoz2014,emdeetal2017}, and that a similarly efficient form 
of high-contrast imager can be utilized simultaneously, it might be conceivable to detect 
Proxima b's polarization either in the continuum or in molecular absorption bands. 
Since the photon rate of polarized photons is less than for the total number of photons, 
the effort of such a search would expectedly be larger than but overall on the same order
as a search in brightness.
}\\

The following summarizes some of the other points discussed earlier:
\begin{itemize}
\item Polarimetry is a valuable complement to photometric and spectroscopic measurements in the 
atmospheric characterization and mapping of exoplanets.

\item The HDSP-CC technique offers a built-in way to separate 
polarization contributions originating with different radial velocities. 
This may prove useful to disentangle the planet polarization from the polarization introduced
by the ISM and the terrestrial atmosphere.
\textcolor{black}{It may also prove useful to remove the effect of systematics provided 
that the systematics signal is either a smooth function of wavelength or 
mimics the stellar spectrum without shifts in wavelength.}

\item The optimal wavelength choice for application of the HDSP-CC technique will depend
on properties of both the planet and the star.

\item Predicting the polarization properties of an exoplanet atmosphere is challenging. 
Even when other techniques (e.g. transmission spectroscopy, brightness phase
curves) have shed some light on the atmospheric structure, the emergent polarization will
be sensitive to weakly constrained properties such as the composition and size of
condensates within about one optical depth from the atmospheric top.
Inversely, polarization measurements will set valuable constraints on such properties.

\item The HDSP-CC technique is less efficient than broadband polarimetry in its use of 
collected photons. The light collected over wavelengths in which
the stellar spectrum is not strongly structured is essentially wasted. 
The expected efficiency can be quantified prior to the observations.

\item If Rayleigh scattering dominates the planet polarization, the best range of 
orbital phases to investigate the planet is closer to $\alpha$$\sim$60--80$^{\circ}$
(where the polarization intensity $Q'$ is maximum)
than to 90$^{\circ}$ (where the planet fractional polarization $Q'$/$A_{\rm{g}}\Phi(\alpha)$
is maximum).

\item The brightness phase curves of giant exoplanets often 
exhibit post-occultation brightness peaks attributed to non-uniform clouds. 
Our model simulations show that for such configurations $Q'$ 
also peaks after occultation. If possible, the observations of such planets, also in polarization, should 
favor post-occultation phases.

\item The ideal observation targets exhibit large brightness and polarization
magnitudes, although both properties are not always compatible because the two 
properties depend on the details of multiple scattering in the atmosphere. 
Brightness phase curves may guide the identification of potential targets to follow up
with polarimetry. 51 Peg b remains a good candidate for both brightness and polarization
measurements in the immediate future.

\item The planet polarization is very sensitive to both the phase angle and wavelength
of the observations. Collecting observations over a broad range of phase angles may 
wash out some of that dependence. Selecting the optimal phase angle to observe, however,
is not trivial. A reasonable (although not failure-proof) approach may be to assume that
Rayleigh scattering will be prominent, thereby favoring phase angles in the
range $\alpha$$\sim$60--80$^{\circ}$.

\end{itemize}

Ultimately, the simulations presented here encourage dedicated observations from current and future
facilities equipped with high-dispersion spectro-polarimeters.

\cleardoublepage

\acknowledgments

I thank Klaus Strassmeier and Ilya Ilyin (Leibniz-Institute for Astrophysics Potsdam, 
Germany) for encouragement and a useful exchange of information at an early stage of the
project. Special thanks go to Luca Fossati 
(Space Research Institute, Austrian Academy of Sciences, Austria) for a thorough 
reading of the manuscript together with helpful and constructive comments.  

\newpage

\cleardoublepage



\end{document}